\documentclass[%
 reprint,
 amsmath,amssymb,
 aps,
]{revtex4-1}
\usepackage{graphicx}% Include figure files
\usepackage{dcolumn}% Align table columns on decimal point
\usepackage{bm}% bold math
\begin{document}

%\preprint{}

\title{Renormalized contact interaction in degenerate unitary Bose gases}
%\title{Renormalized contact potential in strongly interacting Bose gases}
%\title{Radio Frequency Feshbach Resonance with Large Tunability of Scattering Length}
%Controlling Scattering Lengths and Producing Ultra-Cold Molecules with Radio Frequency Fields

%\author{Yijue Ding$^1$, Jose P. D'Incao$^2$ and Chris H. Greene$^1$}
\author{Yijue Ding }
\email[]{ding51@purdue.edu}
\author{ Chris H. Greene}
\email[]{chgreene@purdue.edu}
%\thanks{purdue university}
\affiliation{Department of Physics and Astronomy, Purdue University, West Lafayette, Indiana, USA
}

%$^2$Purdue Quantum Center, Purdue University, West Lafayette, Indiana, USA

%\affiliation{$^1$Department of Physics and Astronomy, Purdue University\\
%$^2$JILA and Department of Physics, University of Colorado, Boulder}
%\author{Yijue Ding} 
%\email[]{ding51@purdue.edu}
%\affiliation{Department of Physics, Purdue University, West Lafayette, Indiana, USA}
%\author{Jos\'e P. D'Incao}
%\affiliation{JILA, University of Colorado and NIST, Boulder, Colorado, USA}
%\affiliation{Department of Physics, University of Colorado, Boulder, Colorado, USA}
%\author{Chris H. Greene}
%\affiliation{Department of Physics, Purdue University, West Lafayette, Indiana, USA}


\date{\today}

\begin{abstract}
%The radio frequency (RF) field is a promising but less developed tool to control cold collisions. From the few-body perspective, 
We renormalize the two-body contact interaction based on the exact solution of two interacting particles in a harmonic trap. This renormalization extends the validity of the contact interaction to large scattering lengths. We apply this renormalized interaction to a degenerate unitary Bose gas to study its stationary properties and elementary excitations using the mean-field theory and the hyperspherical method. Since the scattering length is no longer a relevant length scale at unitarity, universal properties are obtained that depend only on the average particle density. Our treatment shows that the universal relations for the total energy and for the two-body contact are $E/N=12.67\hbar^2\langle n^{2/3}\rangle/2m$ and $C_{2}/N=11.8\langle n^{1/3}\rangle$ respectively. % Our renormalization theory can predict that the universal scaling of energy is $E/N=12.67\hbar^2\langle n^{2/3}\rangle/2m$ and the corresponding two-body contact scale as $C_{2}/N=11.9\langle n^{1/3}\rangle$ at unitarity.
\end{abstract}


\maketitle

\section{Introduction}
Strongly correlated systems near quantum degeneracy exhibit a wide range of intriguing phenomena. Paradigmatic examples include helium superfluidity and the fractional quantum Hall effect. In the atomic physics realm, the ultracold quantum gas, due to its simplicity, purity and high controllability, is an excellent candidate to be used to study strongly correlated systems. The interaction between cold atoms, which is typically characterized by the $s$-wave scattering length, can be readily controlled through Feshbach resonances\cite{chengchinrmp}. 

The Bose Einstein condensate (BEC) is a highly degenerate quantum system in which the interparticle interaction can also be tuned via a magnetic or other types of Feshbach resonance\cite{becfr}. When the scattering length in a BEC is much larger than any length scale of the system, the gas has reached the so-called unitary regime\cite{castin2012}. However, creating a BEC in the strongly interacting regime or even all the way to unitarity is extremely difficult. The major reason is that the three-body recombination rate at zero temperature in dilute gases is proportional to $a^4$\cite{burke1999prl,macek1999prl,braaten2000prl,braaten2006}, which results in a very short lifetime of the strongly interacting Bose gas. This phenomenon contrasts sharply with the strongly interacting Fermi gas, for which the three-body recombination is suppressed by the Pauli exclusion principle\cite{untaryfermi}. Because of the prohibitively high atom loss rate, it has been considered nearly impossible to access the unitary Bose gas adiabatically. However, a nonadiabatic approach to unitarity has been developed by the JILA group\cite{makotynquench}. In their experimental work, they studied the nonequilibrium dynamics of a degenerate unitary Bose gas and observed important universal properties of the system. 

Although a few theories have been proposed to treat the degenerate unitary Bose gas\cite{pethick2002ubgegs,feizhoubec,stoof2011ubgegs,flaviodmc,feizhourg,castin2006pra}, 
%there is no theory so far that can completely describe this system.
no existing theories so far are capable of completely describing this system. Some theories involve complex derivations such as the renormlization group theory\cite{feizhourg} or extensive computations like Monte Carlo simulation\cite{flaviodmc}. Shortly after the JILA experiment on the unitary Bose gas, various theoretical descriptions were proposed in an effort to explain the experimental results, especially the momentum distribution\cite{braatenbec,johnbohnquench,chevy2014,kiraubg,werner2016}. 
 
%  \textbf{In this article, we discuss the degenerate Bose gas with various interactions at zero temperature.} We develop a renormalized mean field theory to address the strongly interacting Bose gases. The renormalization is based on the exact two interacting particle solution in a trap. We compare the ground state properties with the  experiments especially the momentum distribution and the universality. We simulate the dynamics after a interaction quench. Collective excitations such as the breathing mode are discussed. The renormalization is based on the exact two interacting particle solution in a trap. 

In this article, we introduce a renormalized contact potential similar to that in Ref. \cite{stecherrenormal}, to extend the validity of the zero-range potential to the strongly interacting regime. Then we employ this renormalized potential in company with traditional many-body theories to study the degenerate unitary Bose gas at zero temperature. 
%We use a renormalized contact potential similar to that in Ref. \cite{stecherrenormal} which can characterize arbitrary interactions to study the degenerate Bose gases near unitarity. We mainly focus on discussing the ground state properties and dynamics of this system at zero temperature. Throughout the paper we discuss degenerate Bose gases in an isotropic harmonic trap.
%we mainly focus on discussing strongly interacting Bose gases (degenerate Bose gases near unitarity) at zero temperature. 
The structure of this article is as follows: In section II, we elaborate the renormalization procedure and the physical ideas behind it. After that, we apply this renormalized potential to a few many-body theories
%, specifically the mean-field theory and the hyperspherical method
 and show how they are modified with the inclusion of the renormalization. 
%employ this this renormalized potential to different approximation methods to study degenerate Bose gases with arbitrary interactions. 
In section III, we discuss the stationary properties and elementary excitations of a degenerate unitary Bose gas using our renormalization theory, 
%a few important physical quantities of the unitary Bose gas near quantum degeneracy using our renormalization theory
 and compare our results with other theoretical predictions. We particularly focus on a few important physical observables of the system.  Finally, in section IV, we summarize our work and the most significant findings. %discuss some future directions that may improve the theory of unitary Bose gases. 

\section{Theory of renormalization}
Our renormalization is similar to that in a two-component Fermi gas\cite{stecherrenormal}. Here we summarize the procedure and the physical origin of this renormalization. 
%The detailed renormalization procedure is elaborated in Ref.\cite{stecherrenormal}. Here we summarize the physical idea of such renormalization.
For a uniform gas system, when the range of the two-body interaction %(e.g. van der waals range between alkali metal atoms)
is much smaller than both the scattering length $a$ and the average interparticle distance determined by the particle density $n$, the behavior of the system is characterized by the dimensionless parameter $na^3$. This is also equivalent to the dimensionless parameter $k_F a $, where $k_F=(6\pi^2 n)^{1/3}$ is defined for the Bosonic system in a manner akin to the Fermi momentum. Our idea is to design an effective scattering length $a_{eff}$ that can replace the bare scattering length $a$ to describe the properties of the system. In this case, the dimensionless parameter becomes $k_F a_{eff}$. Therefore, there must be a correspondence between $k_F a_{eff}$ and $k_F a$ characterized by a renormalization function $k_F a_{eff}=\zeta(k_F a)$. 
%Thus, the relation between $k_F a_{eff}$ and $k_F a$ is characterized by a renormalization function $k_F a_{eff}=\zeta(k_F a)$. The renormalization procedure is finding the point wise correspondence between $k_F a_{eff}$ and $k_F a$ and construct the renormaliztion function $\zeta$.
The effective scattering length is designed specifically for a system of two interacting particles in a harmonic trap such that it can exactly describe the atomic ground state energy. 

For two particles in a harmonic trap with a circular frequency $\omega_{ho}$ interacting with a regularized pseudo potential
%The idea of such renormalization is to design an effective scattering length $a_{eff}$ and use the corresponding dimensionless parameter $k_F a_{eff}$ to characterize the behavior of the system. To this end, we design a renormalization function such that $k_F a_{eff}=\zeta(k_F a)$. 
%The renormalization procedure is based on the equivalence of $a_{eff}$ and $a$ in describing the energy of two interacting particles in an isotropic harmonic trap. For two particles interacting with a regularized pseudo potential, 
\begin{equation}
V(\textbf{r})=\frac{4\pi\hbar^2 a}{m}\delta(\textbf{r})\frac{\partial}{\partial r}r,
\end{equation}
the Hamiltonian is given by 
\begin{equation}
H_{2b}=-\frac{\hbar^2}{2m}(\nabla_{\textbf{r}_1}^2+ \nabla_{\textbf{r}_2}^2 )+\frac{1}{2} m\omega_{ho}^2 (\textbf{r}_1^2+\textbf{r}_2^2)+V(\textbf{r}_{12}),
\label{h2b}
\end{equation}
where $\textbf{r}_{12}=\textbf{r}_1-\textbf{r}_2$ is the relative coordinate between two particles. 
The wave function is separable in the center of mass motion and the relative motion, that is, $\Psi=\psi_{cm}(\textbf{R}_{cm})\psi_{rel}(\textbf{r}_{12})$, where $\textbf{R}_{cm}=(\textbf{r}_1+\textbf{r}_2)/2$ is the center of mass coordinate.
The exact solution to the corresponding Schr\"odinger equation $H_{2b}\Psi=E_{exact}(a)\Psi$ has been discussed in Ref. \cite{Busch1998}. The eigen-energy can be written as $E_{exact}(a)=E_{cm}+E_{rel}(a)$, where $E_{cm}=(n_{cm}+3/2)\hbar\omega_{ho}$ corresponds to the center of mass motion in the harmonic trap. The eigen-energy for the relative motion satisfies the following condition:
\begin{equation}
\sqrt{2}\frac{\Gamma\left(-E_{rel}(a)/2\hbar\omega_{ho}+3/4\right)}{\Gamma\left(-E_{rel}(a)/2\hbar\omega_{ho}+1/4\right)}=\frac{l_{ho}}{a},
\label{erel}
\end{equation}
where $l_{ho}=\sqrt{\hbar/m\omega_{ho}}$ is the harmonic trap length.

On the other hand, when the two particles interact with the renormalized contact potential given by
\begin{equation}
\tilde{V}(\textbf{r})=\frac{4\pi\hbar^2 a_{eff}}{m}\delta(\textbf{r})=\frac{4\pi\hbar^2 \zeta(k_F a)}{mk_F}\delta(\textbf{r}),
\label{vrenormal}
\end{equation}
%and the corresponding Hamlitonian is given by 
%\begin{equation}
%\tilde{H}_{2b}=-\frac{\hbar^2}{2m}( \nabla_{\vec{r}_1}^2+ \nabla_{\vec{r}_2}^2 )+\frac{1}{2} m\omega_{ho}^2 (r_1^2+r_2^2)+\tilde{V}(\vec{r})
%\end{equation}
%In the framework of mean field approximation,
we assume the total wave function has a Hartree-Fock (HF) expression $\tilde{\Psi}=\psi(\textbf{r}_1)\psi(\textbf{r}_2)$. 
Consequently, the energy expectation value of the system is given by 
\begin{eqnarray}\nonumber
\xi\{\psi\} &= &\int \left[ 2\psi\left(-\frac{\hbar^2}{2m}\nabla^2 + \frac{1}{2}m\omega_{ho}^2r^2\right)\psi   \right. \\  & & \left.  + \frac{4\pi\hbar^2 a_{eff}}{m}\psi^4 \right] d^3\textbf{r}.
\end{eqnarray}
Since there is no Fermi momentum in few body systems, it is natural to replace $k_F$ in Eq. \eqref{vrenormal} by its average value $\langle k_F\rangle=\int [6\pi^2 2\psi(\textbf{r})^2]^{1/3}\psi(\textbf{r})^2 d^3\textbf{r}$ in this two body case. Minimizing $\xi\{\psi\}$ with the normalization constraint $\langle\psi|\psi\rangle=1$ yields the ground state energy that depends on the effective scattering length:
\begin{equation}
E_{HF}(a_{eff})=\xi\{\psi\}|_{\delta\xi/\delta\psi=0}.
\end{equation}
 %and take the variation $\delta\tilde{H}/\delta\psi=0$ under the normalization condition $\langle\psi|\psi\rangle=1$. This leads to a renormalized version of Gross-Petaevskii (GP) equation. The solution to this GP equation yields a single particle wave fucntion $\psi$. Thus, the two-body ground state energy is given by 
%\begin{eqnarray}\nonumber
%E_{HF}(a_{eff})&= &\int \left[ 2\psi\left(-\frac{\hbar^2}{2m}\nabla^2 + \frac{1}{2}m\omega_{ho}^2r^2\right)\psi   \right. \\  & & \left.  + \frac{4\pi\hbar^2 a_{eff}}{m}\psi^4 \right] d^3\vec{r}
%\end{eqnarray}

%With these two different sets of solutions for two interacting particles in a trap, we conduct the renormalization by enforcing the ground state energies from two different approaches equal to each other, 
In order to make the effective scattering length and the bare scattering length equivalent for this trapped two-particle system, we match the HF energy to the exact energy:
\begin{equation}
E_{HF}(a_{eff})=E_{exact}(a).
\label{energyenf}
\end{equation}
Before matching these two energies, we should note that the exact energy has many branches including a molecular branch for $a>0$. Because the HF approximation describes the lowest atomic gas state, which corresponds to the branch with $n_{cm}=0$ and $(1/2)\hbar\omega_{ho}<E_{rel}<(5/2)\hbar\omega_{ho}$, we match the HF energy to the exact energy in this particular branch. 
%We should notice that the solution to Eq. \eqref{erel}  has many branches including a molecular branch for $a>0$. Since we are interested in the ground state of the atomic Bose gas, we should use the lowest atomic gaseous branch, which lies in the range $(1/2)\hbar\omega_{ho}<E_{rel}<(5/2)\hbar\omega_{ho}$, and the lowest eigen energy for the center of mass motion ($n_{cm}=0$), to conduct this renormalization. 
Since Eq. \eqref{energyenf} yields a pointwise correspondence between $\langle k_F\rangle a_{eff}$ and $\langle k_F\rangle a$, we can numerically interpolate the renormalization function $\langle k_F\rangle a_{eff}=\zeta(\langle k_F a \rangle a)$. This interpolation can be excellently fitted by an analytical expression %with which we can interpolate to get a renormalization function $\zeta(x)$. As shown in Ref.\cite{stecherrenormal}, this interpolation can be excellently fitted by an analytical expression:
\begin{equation}
\zeta(x)=0.395-1.138\arctan(0.362-0.994x),
\label{zetafunc}
\end{equation}
which satisfies the asymptotic conditions $\zeta(+\infty)=2.182$, $\zeta(-\infty)=-1.392$ and $\zeta(k_F a)\rightarrow k_F a$ for $|k_F a|\ll 1$. 
%Since this normalization is based on the exact solution of two particles in a trap, we think the effective scattering length $a_{eff}$ can implicitly account for the two-body correlations that is neglected in the mean-field approximation. 

From the renormalization procedure above, we can see that the renormalized contact potential Eq. \eqref{vrenormal} reproduces the exact energy solution for the system of two interacting particles in a trap. 
The next step is to apply such a renormalized contact potential to many body systems. Since the many body Hamiltonian cannot be diagonalized exactly due to the huge number of degrees of freedom, we must make some aggressive but reasonable approximations, as we will discuss in the following subsections.

%Now that we have developed a renormalized contact interaction which yields exact solution for two particles with mean-field approximation, in the following subsections we apply this contact potential to many-body systems with different approxmiation methods. 

\subsection{Mean-field approach}
One natural and intuitive idea is to generalize the HF approximation employed above along with the renormalized interaction to many body systems. Such an approximation for bosons is also called the mean-field approximation. %The typical mean-field approximation exhibits its excellence in addressing weakly interacting systems, but it fails for strongly interacting systems due to the importance of two-body correlations. However, as the two-body correlations are implicitly included in the renormalized potential, we think using mean-field approximation for the many-body wave function will still predict part of important properties of the strongly interacting Bose gas, and we will show the evidence in following sections.

With the inclusion of renormalized interactions, the $N$-body Hamiltonian now becomes
\begin{equation}
H=\sum_{i=1}^N \left(-\frac{\hbar^2}{2m}\nabla_i^2+\frac{1}{2}m\omega_{ho}^2r_i^2 \right)+\sum_{i<j}^N\frac{4\pi \hbar^2 \zeta(k_F a)}{m k_F}\delta(\textbf{r}_{ij}).
\label{hamiltonian}
\end{equation}
The mean-field theory assumes the $N$-body ground state wave function to be $\Psi=\prod_{i=1}^N\psi(\textbf{r}_i)$. By taking the variation $\delta H/\delta\psi=0$ under the normalization condition $\langle\psi|\psi\rangle=1$, we can obtain a renormalized $N$-body Gross-Pitaevskii (GP) equation:
\begin{eqnarray}\nonumber
\left[-\frac{\hbar^2}{2m}\nabla^2+\frac{1}{2}m\omega_{ho}^2r^2  +  \frac{4\pi(N-1)\hbar^2 }{3m} \right. \\ \left. \times \left( \zeta^\prime(k_F a)a+2\frac{\zeta(k_F a) }{k_F} \right) |\psi|^2  \right]\psi=
\epsilon\psi.
\label{gpeq}
\end{eqnarray}
where $\epsilon$ is the Lagrange multiplier enforcing normalization in the variation procedure, and is also identified as the orbital energy. $\zeta'(x)$ means the derivative of $\zeta$ with respect to the variable $x$. $k_F$ is the local Fermi momentum. With such a mean-field approximation and in the framework of the local density approximation (LDA), the local Fermi momentum is given by $k_F=(6\pi^2 N|\psi|^2)^{1/3}$.
%From the same variational procedure in the traditional mean-field theory, we obtain a nonlinear Schr\"odinger equation, 
%\begin{eqnarray}\nonumber
%[H_0+ \frac{4\pi(N-1)\hbar^2 }{3m} \left( 6\pi^2 N |\psi|^2 \right) ^{1/3}\left( \frac{\zeta^\prime(k_F a)a}{k_F} \right. \\ \left. -\frac{\zeta(k_F a) }{k_F^2} \right) |\psi|^2  +  \frac{4\pi (N-1)\hbar^2}{m}\frac{\zeta(k_F a_0)}{k_F} |\psi|^2 ]\psi=
%\epsilon\psi.
%\end{eqnarray}
%\begin{eqnarray}\nonumber
%\left[H_0+ \frac{4\pi(N-1)\hbar^2 }{3m}\left( \zeta^\prime(k_F a)a-\frac{\zeta(k_F a) }{k_F} \right) |\psi|^2 \right. \\ \left. +  \frac{4\pi (N-1)\hbar^2}{m}\frac{\zeta(k_F a_0)}{k_F} |\psi|^2 \right]\psi=
%\epsilon\psi.
%\end{eqnarray}
%where $\epsilon$ is the chemical potential of the system 
%which is also a renormalized version of GP equation. The ground state energy is now given by
%The total energy of the system is given by
After solving Eq. \eqref{gpeq}, we can evaluate the total energy of the system as
\begin{eqnarray}\nonumber
E=\int \left[ N\psi\left(-\frac{\hbar^2}{2m}\nabla^2+\frac{1}{2}m\omega_{ho}^2 r^2\right)\psi \right. \\ \left. +\frac{N(N-1)}{2} \frac{4\pi\hbar^2 \zeta(k_{F}a)}{m k_{F}}\psi^4 \right] d^3\textbf{r}.
\label{egsmf}
\end{eqnarray}

At very large particle numbers, the kinetic energy composes a very small portion of the total energy of the system. In the meanwhile, the particle density varies slowly in the trap except for the edge of the cloud, which is an evidence of the validity of the LDA. In this condition, we can solve Eq. \eqref{gpeq} using the Thomas-Fermi approximation, which neglects the kinetic energy term and thereby converts Eq. \eqref{gpeq} to a regular algebraic equation. In the unitary regime, if we neglect edge effects, which means assuming $k_F a\rightarrow +\infty$ at any position of the cloud, we can obtain an analytical expression for the wave function, which is given by
\begin{equation}
\psi_{TF}(\textbf{r})=\left[ \frac{3(6\pi^2)^{1/3}(R_{TF}^2-r^2)}{16\pi N^{2/3}\zeta(+\infty) l_{ho}^4}  \right]^{3/4},
\label{psitf}
\end{equation}
where $R_{TF}$ is the Thomas-Fermi radius indicating the size of the cloud. It is given by
\begin{equation}
R_{TF}=N^{1/6}\left(\frac{256\sqrt{2}}{9}\right)^{1/6}\left(\frac{\zeta(+\infty)}{\pi}\right)^{1/4} l_{ho}.
\end{equation}
Consequently, the orbital energy is given by 
\begin{equation}
\epsilon=\frac{1}{2}m\omega_{ho}^2R_{TF}^2.
\end{equation}
The fact that $\psi_{TF}$ does not depend on $a$ is another signature that the scattering length is no longer a relevant length scale in the unitary regime. 

In order to calculate the dynamics of a degenerate Bose gas, it is natural to convert Eq. \eqref{gpeq} to a time-dependent GP equation: 
\begin{eqnarray}\nonumber
\left[-\frac{\hbar^2}{2m}\nabla^2+\frac{1}{2}m\omega_{ho}^2r^2  +  \frac{4\pi(N-1)\hbar^2 }{3m} \right. \\ \left. \times \left( \zeta^\prime(k_F a)a+2\frac{\zeta(k_F a) }{k_F} \right) |\tilde{\psi}|^2  \right]\tilde{\psi}=
i\hbar\frac{\partial}{\partial t}\tilde{\psi}.
\label{gpeqtd}
\end{eqnarray}
We should notice that now the local Fermi momentum $k_F$ also becomes time-dependent. Although such a non-linear and time-dependent Shr\"odinger equation can be solved directly using a brute force time-evolution method, we can obtain a clearer physical picture if appropriate approximations are made to Eq. \eqref{gpeqtd}. 
%clear physics can be obtained with some approximations to linearize Eq. \eqref{gpeqtd}. 
One typical method is the Bogoliubov approximation, which is commonly used to predict elementary excitations of a BEC. 
%The elementary excitations can be predicted with the Bogoliubov approximation, which assumes the time dependent wave function to be
The Bogoliubov approximation assumes the time-dependent wave function to be
\begin{equation}
%\psi_{bog}(\vec{r},t)=e^{-i\epsilon t/\hbar}\left[\psi(\vec{r}) + \sum_\gamma\left(u_\gamma (\vec{r})e^{-i\omega_{\gamma}t}+v_\gamma^*(\vec{r})e^{i\omega_{\gamma}t}\right) \right].
\tilde{\psi}_{bog}(\textbf{r},t)=e^{-i\epsilon t/\hbar}\left(\psi(\textbf{r}) + u(\textbf{r})e^{-i\omega t}+v^*(\textbf{r})e^{i\omega t} \right).
\end{equation}
Inserting this wave function into Eq. \eqref{gpeqtd} and linearizing the equation to the first order in $u(\textbf{r})$ and $v(\textbf{r})$, we can obtain a pair of coupled differential equations:
\begin{eqnarray}\nonumber
\left(-\frac{\hbar^2}{2m}\nabla^2+\frac{1}{2}m\omega_{ho}^2r^2 + f(N, a, \psi)-\epsilon \right)u \\+g(N,a,\psi)v=\hbar\omega u,
\end{eqnarray}
\begin{eqnarray}\nonumber
\left(-\frac{\hbar^2}{2m}\nabla^2+\frac{1}{2}m\omega_{ho}^2r^2 + f(N, a, \psi)-\epsilon \right)v \\+g(N,a,\psi)u=-\hbar\omega v,
\end{eqnarray}
where $\omega$ corresponds to the eigen mode frequency. $f$ and $g$ are in general complicated functions of $N$, $a$, and $\psi$, but they have simple forms in the asymptotic limits $a\rightarrow 0$ and $a\rightarrow\infty$:
\begin{equation}
f\rightarrow \frac{8\pi N\hbar^2a}{m}\psi^2, g\rightarrow -\frac{4\pi N\hbar^2a}{m}\psi^2 \quad (a\rightarrow 0),
\end{equation}
and 
\begin{eqnarray}\nonumber
f\rightarrow \frac{40\pi N^{2/3}\hbar^2\zeta(+\infty)}{9(6\pi^2 )^{1/3}m}\psi^{4/3},  & \\ g\rightarrow -\frac{16\pi N^{2/3}\hbar^2\zeta(+\infty)}{9(6\pi^2 )^{1/3}m}\psi^{4/3}  & \quad (a\rightarrow\infty).
\end{eqnarray}
The lowest and most significant eigen-mode is called the breathing mode, which corresponds to the oscillation of the overall size of the cloud with a fixed geometry. We will discuss this mode in later sections. 
\subsection{Hyperspherical description}
The hyperspherical coordinate system is a powerful toolkit to treat few-body problems\cite{javiernatphys,wangjia2012,wangyujun2012,sethhyper}. To generalize this toolkit to many body systems, aggressive approximations must be made to significantly reduce the dimension of the problem.
%since the many body Hamiltonian are unable to diagonalized exactly due to the huge degrees of freedom. 
The hyperspherical description of a single component weakly-interacting BEC and a degenerate Fermi gas have been studied with the bare Fermi pseudo potential, and important physics has been predicted even with crude approximations\cite{bohn1998,sethpra2006}. Similar to the procedure in Ref. \cite{bohn1998}, we formulate the hyperspherical theory in a degenerate unitary Bose gas with the renormalized interaction.%we formulate the hyperspherical method with our renormalized contact potential to generalized the theory to strongly interacting regime. 

For $N$ particles in a 3D space, which contains $3N$ degrees of freedom, the hyperspherical coordinates are constructed as follows: The hyperradius $R$, which is a collective coordinate and indicates the overall size of the cloud, is given by
\begin{equation}
R=\sqrt{\frac{1}{N}\sum_{i=1}^N r_i^2},
\end{equation}
The remaining $3N-1$ coordinates are called hyperangles. $2N$ of them are the defined as the regular spherical angles of the $N$ particles, that is, $\{\theta_1,\phi_1,\theta_2,\phi_2,\cdots,\theta_N,\phi_N\}$. The remaining $N-1$ hyperangles can be defined as 
\begin{equation}
\tan\alpha_i=\frac{\sqrt{\sum_{j=1}^i r_j^2}}{r_{i+1}}, (i=1,\cdots,N-1)
\end{equation}

With this set of hyperspherical coordinates, the Hamiltonian in Eq.\eqref{hamiltonian} can be rewritten as
\begin{eqnarray}\nonumber
%H=-\frac{\hbar^2}{2M}\frac{1}{R^{(3N-1)/2}}\frac{\partial^2}{\partial R^2}R^{(3N-1)/2}+\frac{(3N-1)(3N-3)\hbar^2}{2}
H &= &-\frac{\hbar^2}{2M}\frac{1}{R^{3N-1}}\frac{\partial}{\partial R}R^{3N-1}\frac{\partial}{\partial R} +\frac{\Lambda^2}{2MR^2} \\ & &  +\frac{1}{2}M\omega_{ho}^2R^2+V_{int}(R,\Omega),
\end{eqnarray}
where $M=Nm$ is the total mass of the $N$ particles. $V_{int}(R,\Omega)$ denotes the renormalized interactions written in hyperspherical coordinates, that is,
\begin{equation}
V_{int}(R,\Omega)=\sum_{i<j}^N\frac{4\pi \hbar^2 \zeta(k_F a)}{m k_F}\delta(\textbf{r}_{ij}).
\end{equation}
$\Lambda$ is called the grand angular momentum operator. The eigen-functions of the operator $\Lambda^2$, denoted by $\Phi_\lambda$, are called hyperspherical harmonics\cite{hyperspherical}. They satisfy the equation
\begin{equation}
\Lambda^2 \Phi_\lambda(\Omega)=\lambda(\lambda+3N-2)\Phi_\lambda(\Omega),
\end{equation}
where $\Omega$ represents all hyperangles. For a given $\lambda$, $\Phi_\lambda$ usually has huge degeneracy especially for large $\lambda$. 
The eigen-functions $\Phi_\lambda$ form a basis in the hyperangular Hilbert space. An aggressive approximation we make here is to only retain one hyperangular momentum eigen-state out of this huge basis set. This is also known as the $K$-harmonics approximation in nuclear theories\cite{smirnov1977} and it becomes exact in the unitary limit and in the non-interacting limit\cite{castin2006pra,blume2008pra}. The natural choice of this eigen state for bosons would be the lowest eigen state of $\Lambda^2$, denoted by $\Phi_0$, which in fact is a constant. Such a choice of hyperangular wave function also freezes the geometry of the atomic cloud into that of the non-interacting case,
%denotes that the geometry of the atomic cloud is frozen to be the non-interacting case, 
while the interparticle interactions modify the overall size of the cloud, which is reflected in the hyperradial wave function.
With this $K$-harmonics approximation, the total $N$-body wave function can be separated as $\Psi(R,\Omega)=F(R)\Phi_0(\Omega)$. Inserting this expression into the time-independent Schr\"odinger equation and integrating over all hyperangles yields a hyperradial Schr\"odinger equation:
%The $N$-body wave function is given by 
%\begin{equation}
%\Psi(\vec{r}_1,\vec{r}_2,\cdots , \vec{r}_N)=F(R)\Phi_{\lambda}(\Omega)
%\end{equation}
%where $R$ is the hyperradius, $\omega$ includes the $N-1$ hyper angles, $\Phi_\lambda$ is the eigen function of the operator $\Lambda^2$, which is called hyperspherical harmonics\cite{hyperspherical}.
%\begin{equation}
%\Lambda^2 \Phi_\lambda(\omega)=\lambda(\lambda+3N-1)\Phi_\lambda(\Omega)
%\end{equation}
%The pseudo potential coefficient before renormalization is $g=4\pi\hbar^2 a/m$, we choose the non-interacting Bose gas ground state energy $E_{NI}=3N\hbar\omega/2$ as an energy scale and the corresponding average hyperradius (at large $N$ limit) $R_{NI}=\sqrt{3/2}l_{ho}$ as an length scale. $l_{ho}=\sqrt{\hbar/m\omega}$ is the harmonic oscillator length. 
%The hyperradial Schr\"odinger equation is given by
%\begin{equation}
%\left(-\frac{\hbar^2}{2M}\frac{d^2}{dR^2}+U_{eff}(R)\right)R^{(3N-1)/2}F(R)=ER^{(3N-1)/2}F(R)
%\end{equation}
\begin{equation}
\left(-\frac{\hbar^2}{2M}\frac{d^2}{dR^2}+V_{eff}(R)-E\right)R^{(3N-1)/2}F(R)=0,
\label{hyperradialeq}
\end{equation}
where $V_{eff}(R)$ is an effective hyperradial potential written as
\begin{eqnarray}\nonumber
%V_{eff}(R)=\frac{(3N-1)(3N-3)\hbar^2}{8MR^2}+\frac{1}{2}M\omega^2 R^2 \\ +\langle\Phi_\lambda(\Omega)|\sum_{i<j}^N U_{int}(\vec{r}_{ij}) |\Phi_\lambda(\Omega)\rangle_\Omega
V_{eff}(R)=\frac{(3N-1)(3N-3)\hbar^2}{8MR^2}+\frac{1}{2}M\omega_{ho}^2 R^2 \\ +\langle\Phi_0(\Omega)|V_{int}(R,\Omega)|\Phi_0(\Omega)\rangle_\Omega,
\end{eqnarray}
where $\langle\cdots\rangle_\Omega$ denotes integration over all hyperangles. % and $\vec{r}_{ij}=\vec{r}_i-\vec{r}_j$
The evaluation of the last term in $V_{eff}$ has been elaborated in Ref. \cite{bohn1998}. In the large $N$ limit which we are interested in, $\langle\Phi_0|V_{int}(R,\Omega)|\Phi_0\rangle_\Omega$ can be readily calculated numerically. We can even obtain analytical expressions in the unitary limit and the weakly interacting limit: For $a\rightarrow\infty$, % and in $a\rightarrow 0$ and $a\rightarrow\infty$ conditions. It can be evaluated analytically as
\begin{equation}
\langle\Phi_0|V_{int}(R,\Omega)|\Phi_0\rangle_\Omega=\frac{N^{8/3}(3/5)^{3/2}\zeta(+\infty)(4\pi/3)^{1/3}\hbar^2}{\pi M R^2},
\end{equation}
and for $a\rightarrow 0$, 
\begin{equation}
\langle\Phi_0|V_{int}(R,\Omega)|\Phi_0\rangle_\Omega=\frac{N^3(2/\pi)^{1/2}(3/2)^{3/2}\hbar^2a}{2MR^3}.%\frac{N^3}{2}\sqrt{\frac{2}{\pi}}\left(\frac{3}{2}\right)^{3/2}\frac{\hbar^2 a}{MR^3}
\end{equation}

\begin{figure}
\includegraphics[width=0.48\textwidth]{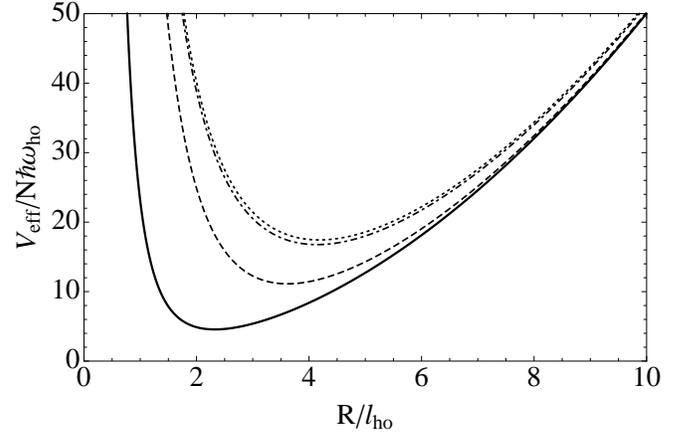}
\caption{Effective hyperradial potential curves for $N=5000$ particles with different scattering lengths $a/l_{ho}=$0.01(solid), 0.1(dashed), 1(dotdashed) and 10 (dotted) respectively.}
\label{fig1}
\end{figure}

Fig. 1 shows the effective hyperradial potential curves at different scattering lengths. At large hyperradius, $V_{eff}(R)$ is always dominated by the $R^2$ term representing the confinement of the harmonic trap. At a small hyperradius, the system feels two repulsive forces originating from the quantum pressure and the two-body interaction. It is interesting that the two-body interaction term transitions from $R^{-3}$ to $R^{-2}$ as the scattering length increases to infinity. 
%At the long range limit, the effective potential scale as $R^2$, which denote the cloud is confined by the harmonic trap. In the short range, the effective potential has a repulsive wall which scales as $R^{-2}$, even without interaction, which represents the quantum pressure induced by the kinetic energy. at small interactions, the repulsive wall scales as $R^{-3}$, which means the repulsive interaction between Bosons is the major contribution that prevents the cloud to collapse.  What’s intriguing is that at unitary limit, the short range repulsive potential scale as $R^{-2}$ again.

Eq. \eqref{hyperradialeq} is equivalent to the Schr\"odinger equation of a particle moving in a 1D potential. Moreover, in the large $N$ limit, the mass of this ``particle" is so huge that it can be treated classically. Consequently, 
the minimum of $V_{eff}$ corresponds to the total energy of the system at equilibrium, that is,
\begin{equation}
E=V_{eff}(R_0),
\end{equation}
where $R_0$ denotes the equilibrium position. 
Furthermore, as the hyperangular wave function is kept frozen, the oscillation of this massive ``particle" in the hyperradial potential indicates the oscillation of the overall size of the system, which corresponds to the breathing mode. The breathing mode frequency is associated with the coefficient of the second order expansion of $V_{eff}$ at $R_0$, that is, % since the hyperradius denotes the overall size of the cloud while the hyperanglar wave function is kept frozen. This denotes the breathing mode of the system, 
%taking the series expansion of the effective hyperradial potential at the minimum position, The second order coefficient indicates the breathing mode frequency, which is given by 
\begin{equation}
\omega=\omega_{ho}\sqrt{\frac{1}{M}\left. \frac{d^2 V_{eff}}{dR^2}\right\vert_{R=R_0}}%\sqrt{\frac{1}{m}\left[\frac{d^2}{dR^2}\left(\frac{U_{eff}}{N}\right)_{R=R_{min}}\right]}
\end{equation}
%This means the minimum of $V_{eff}$ represents the ground state energy and the corresponding hyperradius is the euilibrium point of the cloud.
%For $N\gg 1$ and $a\rightarrow\infty$, the equilibrium position is given by 
%\begin{equation}
%R_{min}=N^{1/6}(3/5)^{3/8}[2\zeta(+\infty)/\pi]^{1/4}(4\pi/3)^{1/12}l_{ho}%N^{1/6}\left(\frac{3}{5}\right)^{3/8}\left(\frac{4\pi}{3}\right)^{1/12}\left(\frac{2\zeta(+\infty)}{\pi}\right)^{1/4}l_{ho}
%\end{equation}
%and the ground state energy per particle is given by 
%\begin{equation}
%\frac{E_{gs}}{N}=\hbar\omega_{ho}N^{1/3}(3/5)^{3/4}[2\zeta(+\infty)/\pi]^{1/2}(4\pi/3)^{1/6}
%\end{equation}
%\begin{eqnarray}\nonumber
%\label{meanenergy}
%\frac{V_{eff}}{E_{NI}} &=&\frac{(3N-1)(3N-3)}{18N^2(R/R_{NI})^2} \\&& +\frac{1}{2}(R/R_{NI})^2+\frac{(N-1)(a/R_{NI})}{\sqrt{3\pi}(R/R_{NI})^3}
%\end{eqnarray}

\section{Results and discussions}
%Equipped with this renormalized mean-field toolkit, we can solve the mean-field wave function at any interaction strength. Fig. shows the ground state wave functions of BEC with $a=142$au and the unitary Bose gas respectively\cite{radzihovskybec}. At unitarity, particles strongly push against each other, expanding the atomic cloud in the trap. For weakly interacting BEC, the interaction parameter is $(N-1)a/l_{ho}$
%The next calculation is given by 
%The BEC ground state energy with LHY correction\cite{LHYcorrection,becdilutegas} in an external potential, with mean field wave function $n(\vec{r})=N|\psi(\vec{r})|^2$, is given by
%\begin{eqnarray}\nonumber
%E_{gs}=\int\left\lbrace \frac{n^2(\vec{r})U_0}{2}\left[1+\frac{128}{15\sqrt{\pi}}(n(\vec{r})a^3)^{1/2}\right] \right. \\ \left. +\frac{\hbar^2}{2m}|\nabla \sqrt{n(\vec{r})}|^2+n(\vec{r})V_{ext}(\vec{r})\right\rbrace dV
%\end{eqnarray}
%\subsection{Ground state energy and condensate fraction}
%We calculate the ground state wave function and plot the momentum distribution and compare it with the experiment\cite{johnbohnquench}. The two-body and three-body contact is given by Ref.\cite{braatencontact,braatenbec}.

%=================================Ground state energy discussion===============================================
At first, we discuss the total energy of a degenerate Bose gas. Fig. 2 shows the average energy per particle for $N=10^4$ particles as a function of the scattering length. In the weakly interacting regime where $\langle n \rangle a^3\ll 1$, the mean-field energy obtained by solving the GP equation with renormalization agrees excellently with that without renormalization. These two results start to separate near $a/l_{ho}=0.2$, which corresponds to $\langle n\rangle a^3=0.25$. The mean-field energy diverges at unitarity without renormalization. The energy obtained using the Thomas-Fermi approximation and the mean-field energy differ in the weakly interacting regime where the kinetic energy is a significant contribution to the total energy. However, at large scattering lengths, the total energy of the system is dominated by the strong interactions between particles and thereby the Thomas-Fermi approximation agrees excellently with the mean-field result. The energy obtained using the hyperspherical method agrees qualitatively with the mean-field result, though it is slightly smaller. Overall, the total energy of the system saturates at large scattering lengths with the inclusion of the interaction renormalization, which indicates that the scattering length is no longer a relevant length scale of the system near unitarity.

\begin{figure}
\includegraphics[width=0.48\textwidth]{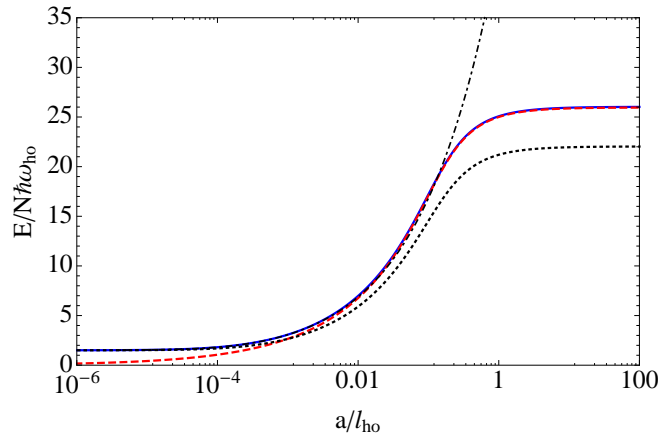}
\caption{The average energy per particle as a function of the scattering length in the ground state of a degenerate Bose gas for $N=10^4$ particles. The results are obtained with the renormalized interaction using mean-field approach(blue solid), the Thomas-Fermi approximation(red dashed), and hyperspherical method(black dotted) respectively. The dotdashed curve shows the mean-field result with the bare unrenormalized contact interaction for comparison. }
%(a) Energy per particle in the ground state as a function of scattering with the renormalized contact potential using mean-field method (solid ), Thomas-Fermi approximation(dashed)  and K-harmonic approximation(dotted). The number of particle is $N=10^3$, (b) the same as (a) except for $N=10^5$.
\label{fig2}
\end{figure}

%From Figure 2, we know that Thomas-Fermi approximation is very accurate when the interaction is strong, 
We now discuss the energy of a unitary Bose gas using the Thomas-Fermi approximation since it is very accurate in the strongly interacting regime. 
One advantage of the Thomas-Fermi approximation is that we can obtain analytical expressions for many physical quantities, which offers us a clear picture of the unitary Bose gas. 
%The Thomas-Fermi offers us analytical solutions almost for everything. The analytical solution of the wave function Eq. \eqref{psitf} offers us a clear picture of the unitary Bose gas. 
With the Thomas-Fermi approximation, the ground state energy is given by
\begin{equation}
\frac{E}{N}=\frac{27}{64}\left(\frac{256\sqrt{2}}{9}\right)^{1/3}\left(\frac{\zeta(+\infty)}{\pi}\right)^{1/2}N^{1/3}\hbar\omega_{ho}
\label{eovern}
\end{equation}

For a unitary gas in a uniform space, the only relevant length scale of the system is the average interparticle distance determined by $n^{-1/3}$, where $n$ is the particle density. This also defines the only energy scale of the system $\hbar^2n^{2/3}/2m$. When the gas is inhomogeneous while the density varies slowly in space, the local density approximation can be applied to the system and thereby $n^{2/3}$ is replaced by its average value $\langle n^{2/3}\rangle$. %This is valid when the density varies slowly in the trap. %the ground state energy per particle should scale as $n^{2/3}$, 
For a unitary Bose gas in a harmonic trap, the average value is given by
\begin{equation}
\langle n^{2/3}\rangle=\frac{5\times 3^{2/3}N^{1/3}}{8(2\pi)^{5/6}\zeta(+\infty)^{1/2}l_{ho}^2}.%\frac{N^{1/3}}{l_{ho}^2}
\label{density23}
\end{equation}
Therefore, from Eq. \eqref{eovern} and Eq. \eqref{density23}, we can obtain the universal relation of the energy of a unitary Bose gas, which is given by
%The relation between energy and the density is 
\begin{equation}
\frac{E}{N}=\frac{6^{5/3}\pi^{1/3}\zeta(+\infty)}{5}\frac{\hbar^2\langle n^{2/3}\rangle}{2m}\approx 12.67\frac{\hbar^2\langle n^{2/3}\rangle}{2m}
\label{euniversal}
\end{equation}
This universal relation is close to the value $E/N=13.33\hbar^2 n^{2/3}/2m$ reported in Ref. \cite{pethick2002ubgegs}.

\begin{figure}
\includegraphics[width=0.48\textwidth]{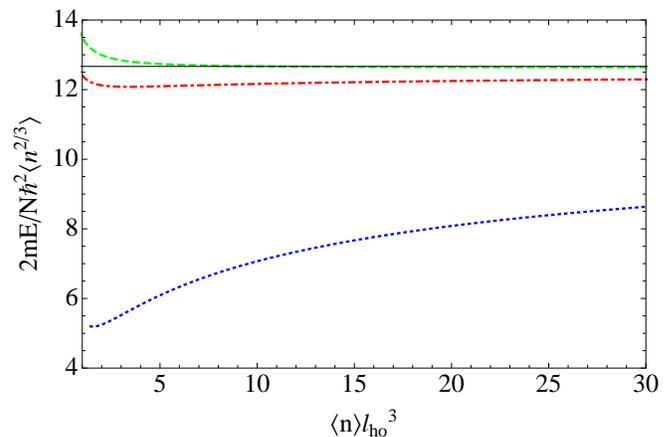}
\caption{The average energy per particle, in units of $\hbar^2\langle n^{2/3}\rangle/2m$, as a function of the average particle density for $a/l_{ho}=0.1$(blue dotted), $1$(red dotdashed) and $10$(green dashed) respectively. The black solid line shows the universal relation of energy from Eq. \eqref{euniversal}.}
\label{fig3}
\end{figure}
We show in Fig. 3 the average energy per particle, in units of $\hbar^2\langle n^{2/3}\rangle/2m$, as a function of the average density for different scattering lengths to verify this universal relation. The results are obtained by solving the renormalized GP equation Eq. \eqref{gpeq}. At small scattering length $a/l_{ho}=0.1$, the value of $2mE/N\hbar^2\langle n^{2/3}\rangle$ varies significantly with the average density. As the scattering length increases, $2mE/N\hbar^2\langle n^{2/3}\rangle$ has weaker dependence on the average density and the value approaches the universal constant in Eq. \eqref{euniversal}. At $a/l_{ho}=10$, where gas has reached the unitary regime, the result agrees excellently with the universal relation Eq. \eqref{euniversal} for $\langle n\rangle l_{ho}^3>5$, The small deviation from the universality at small densities may be due to the inaccuracy of LDA when the interparticle distance is comparable to the trapping length. 
%The universal relation of energy that we predicted from our renormalization theory is very close to the value/similar to that reported by Ref. \cite{pethick2002ubgegs,feizhoubec,stoof2011ubgegs}.
%The energy per particle in the ground state is shown in Figure 2 with different calculation methods. 
%The condensate fraction is shown in Figure 3.
%The effect of three-body loss and interactions on the momentum distribution of a unitary Bose gas\cite{chevy2014}

%=================================Two-body contact discussion===============================================
%Next, as we discuss the Tan's contact near unitarity.
Besides the total energy, there are many interesting physical quantities worthy of investigation in unitary Bose gases. 
An important quantity that bridges the two-body correlations and the thermodynamics of a many-body system is called the two-body contact or Tan's contact. It was first introduced by Shina Tan to study the universal properties of a two-component Fermi gas with $s$-wave contact interactions\cite{tancontact1,tancontact2}. Universal relations determined by the two-body contact have also been identified in systems consisting of identical bosons\cite{braatencontact}. The two-body contact in bosons has been measured using rf spectroscopy\cite{jincontact2012}.
The two-body contact is determined by the derivative of the total energy of the system with respect to the scattering length: 
\begin{equation}
C_2= \frac{8\pi m a^2}{\hbar^2}\frac{d E}{d a}.
\end{equation}
It is an extensive thermodynamic quantity of the system. Another intrinsic quantity, which is commonly used in homogeneous systems, is the contact density $\mathcal{C}_2$, %which has the dimension of $k^4$, 
which can be obtained from the limit of the high momentum tail: $\mathcal{C}_2=\lim_{k\rightarrow\infty}k^4n_\textbf{k}$, where $n_\textbf{k}$ is the number of particles in the $\textbf{k}$ momentum state. Since the interparticle distance is the only relevant length scale of a homogeneous system at unitarity, the two-body contact density must scale as
\begin{equation}
\mathcal{C}_2=\alpha n^{4/3},
\end{equation}
where $\alpha$ is an universal dimensionless coefficient. Such a universal relation can be generalized to a trapped system under LDA, which is given by
\begin{equation}
C_2=\alpha N\langle n^{1/3}\rangle.
\label{c2universal}
\end{equation}

\begin{figure}
\includegraphics[width=0.48\textwidth]{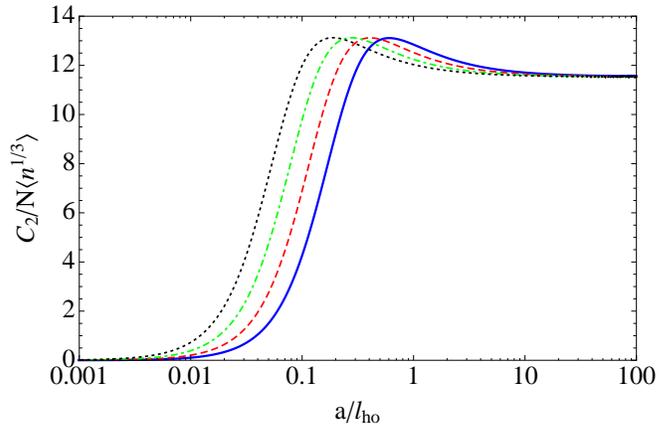}
\caption{The average two-body contact per particle, in units of $\langle n^{1/3}\rangle$, as a function of the scattering length for $N=10^3$(blue solid), $10^4$(red dashed),$10^5$(green dotdashed) and $10^6$(black dotted) particles.}%and $10^5$(black)
%\caption{Breathing mode frequency as a function of the scattering length for $N=10^3$(blue, solid) and $N=10^5$(red, dashed) with K-harmonic approximation.}
\label{fig4}
\end{figure}
Fig. 4 shows the average two-body contact per particle, in units of $\langle n^{1/3}\rangle$, as a function of the scattering length. The results are obtained by directly solving the renormalized GP equation Eq. \eqref{gpeq}.
The value of $C_2/(N\langle n^{1/3}\rangle)$ has similar behavior for different numbers of particles: 
It increases drastically in the weakly interacting regime as the scattering length grows; it then attains a maximum value before saturating in the unitary regime.
%With the increment of the scattering lengths, it increases drastically in the weakly interacting regime; then it attains a maximum value before it saturates in the unitary regime. % then it saturates at the same value at large scattering lengths for different number of particles. 
For different numbers of particles $C_2/(N\langle n^{1/3}\rangle)$ saturates at the same value, demonstrating the universality of $\alpha$.
\begin{figure}%[h]
\includegraphics[width=0.48\textwidth]{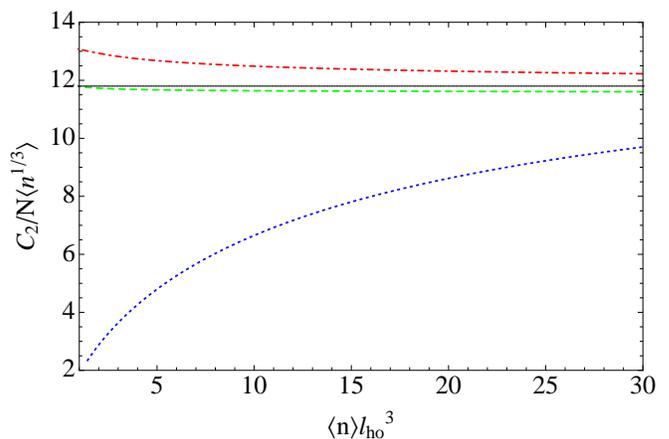}
\caption{The average two-body contact per particle, in units of $\langle n^{1/3}\rangle$, as a function of the average particle density for $a/l_{ho}=0.1$(blue dotted), $1$(red dotdashed) and $10$(green dashed) respectively. The black solid line shows the universal relation of two-body contact from Eq. \eqref{c2universal}.}
%\caption{Two-body contact in units of $n^{4/3}$ as function of the scattering length. The number of particle is $N=10^5$}
\label{fig5}
\end{figure}

From our renormalized mean-field model, we can derive the universal relation Eq. \eqref{c2universal} and determine the universal coefficient $\alpha$ analytically with appropriate approximations. 
To calculate the two-body contact, we take the derivative of the renormalized mean-field energy Eq. \eqref{egsmf} with respect to the scattering length. 
We should note that at unitarity %the dependence of $\psi$ on the scattering length is much smaller than the renormalization function $\zeta(k_F a)$. 
the scattering length dependence of the wave function $\psi$ is much weaker than the renormalization function $\zeta(k_F a)$. 
Thus, by approximating $\partial\psi/\partial a|_{a\rightarrow\infty}\approx 0$ and neglecting the edge effect, we can readily obtain Eq. \eqref{c2universal} and the coefficient $\alpha$ is given by
%Taking the derivative of the renormalized mean-field energy Eq. \eqref{egsmf} with respect to the scattering length, while noting that at unitarity the dependence of $\psi$ on the scattering length is much smaller than the renormalization function $\zeta(k_F a)$, we approximate that $\partial\psi/\partial a|_{a\rightarrow\infty}\approx 0$ and neglect the edge effect, we can readily obtain Eq. \eqref{c2universal} and the coefficient $\alpha$ is given by, In our model with the renormalized interaction, the scaling coefficient $\alpha$ is given by
\begin{equation}
\alpha=\frac{1.138(4\pi)^2}{0.994(6\pi^2)^{2/3}}\approx 11.8.
\label{alphavalue}
\end{equation}
Other theoretical works have reported the the universal coefficient to be $\alpha=10.3$\cite{stoof2011ubgegs},  $9.04$\cite{flaviodmc} and $12$\cite{johnbohnquench}, which agree qualitatively with our prediction from the renormalized mean-field approach. 
%The value of the universal coefficient $\alpha$ from our renormalization theory agree qualitatively with other theoretical predictions, which which is similar to the values reported in Ref. \cite{stoof2011ubgegs,johnbohnquench,flaviodmc}. They reported this universal coefficent to be $\alpha= 9.04$\cite{flaviodmc}, $12$\cite{johnbohnquench} and $10.3$\cite{stoof2011ubgegs} .The value is very close to that obtained with far more complicated theories.

To further verify our prediction of the universal relation Eq. \eqref{c2universal} and the value of the coefficient $\alpha$, we show in Fig. 5 the average two-body contact per particle, in units of $\langle n^{1/3}\rangle$, as a function of the average particle density for different scattering lengths.
%The results are obtained by directly solving the renormalized GP equation Eq. \eqref{gpeq} and evaluating the energy derivative with respect to the scattering length. 
At small scattering length $a/l_{ho}=0.1$, $C_2/(N\langle n^{1/3}\rangle)$ increases significantly with the density. The value of $C_2/(N\langle n^{1/3}\rangle)$ at $a/l_{ho}=1$ is larger than its values for both $a/l_{ho}=0.1$ case and $a/l_{ho}=10$ case because it attains a maximum with the increment of the scattering length, as shown in Fig. 4. 
%For the largest scattering length $a/l_{ho}=10$, which represents that the gas is in the unitary regime, 
In the unitary regime,
$C_2/(N\langle n^{1/3}\rangle)$ converges to the universal coefficient $\alpha$, which corresponds to $a/l_{ho}=10$ case. 
The value of $\alpha$ from exact calculation is slightly smaller than the the analytical approximation $\alpha\approx 11.8$, %approximated value $11.8$, 
which might be due to the edge effect and the small scattering length dependence of the wave function. 

%=================================Breathing mode discussion===============================================
\begin{figure}%[h]
\includegraphics[width=0.48\textwidth]{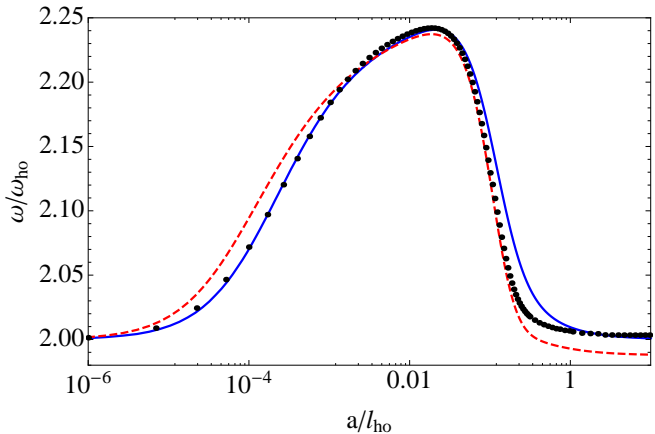}
\caption{The breathing mode frequency as a function of the scattering length for $N=10^4$ particles. The results are obtained using the hyperspherical method(blue solid), Bogoliubov approximation(red dashed) and by direct time evolution(black dots).}
\label{fig6}
\end{figure}
Finally, we discuss the elementary excitations of the degenerate Bose gas in a harmonic trap. Specifically, we focus on the lowest radial excitation, which corresponds to the breathing mode. 
%the unitary Bose gas using different theories with the renormalization. The lowest radial excitation in a trap corresponds to the breathing mode.
The breathing mode frequency can be determined from the hyperspherical method, the Bogoliubov approximation, or by directly solving the time-dependent GP equation, as discussed in the theory of renormalization section above. 
Fig. 6 shows the breathing mode frequency as a function of the scattering length for $N=10^4$ particles. These three different methods show overall consistency with each other. The time-evolution results are obtained by solving Eq. \eqref{gpeqtd} with an interaction quench. At vanishing scattering length, the breathing mode frequency is apparently $\omega=2\omega_{ho}$, which corresponds to the beating between two adjacent harmonic levels with zero angular momentum. The breathing mode frequency increases gradually with the scattering length before it reaches a maximum value. It is interesting that the breathing mode frequency regresses to the non-interacting value $\omega=2\omega_{ho}$ at unitarity. Similar unitary behavior has also been predicted in Ref. \cite{castin2006pra}.  
% It is calculated from $K-$harmonic approximation. People quench the BEC scattering length in the small interaction region and studied the evolution of condensation fraction\cite{hadzibabic2012}. Salomon et al measured the LHY correction and deduced the lower bound of the energy coefficient of a unitary Bose gas\cite{salomon2011}.

%\begin{figure*}
%\label{momentumdist}
%\begin{center}
%\includegraphics[width=\textwidth]{momentum2.eps}
%\caption{(a) The ground state wave function of the BEC and the unitary Bose gas (b)Transverse momentum distribution of the unitary Bose gas immediately after quenching}
%\end{center}
%\end{figure*}

%\begin{figure*}
%\label{propagation}
%\begin{center}
%\includegraphics[width=\textwidth]{propagation.eps}
%\caption{Evolution of BEC cloud after quenching to unitarity. (a) $ t=0$. (b)$ t=0.5\pi/\omega_{ho}$. (c) $t=\pi/\omega_{ho}$.}
%\end{center}
%\end{figure*}

\section{Conclusion}
In summary, we introduced a renormalized contact potential to study degenerate Bose gases with large scattering lengths. Such a renormalized interaction is designed by matching the Hartree-Fock energy to the exact energy of two interacting particles in a trap. We employ this renormalized contact potential in company with the mean-field theory and the hyperspherical theory to study the stationary properties and elementary excitations of a degenerate Bose gas, especially in the unitary regime. %For the stationary properties, this renormalization offers us a much more clear and convenient approach to obtain the universal relations for energy and two body contact at unitarity. The results are in consistent with other theoretical predictions. 
In the framework of the local density approximation, the only relevant length scale of a degenerate unitary Bose gas is the interparticle spacing $n^{-1/3}$, where $n$ is the particle density. This length scale also determines the only energy scale $\hbar^2 n^{2/3}/2m$ and the only two-body contact scale $n^{1/3}$ of the system. Our renormalization theory offers us a much more clear and convenient approach to obtain the universal relations for energy and two-body contact at unitarity, which are given by $E/N=12.67\hbar^2\langle n^{2/3}\rangle/2m$ and $C_{2}/N=11.8\langle n^{1/3}\rangle$ respectively. Our results are in consistent with other theoretical predictions. 
Moreover, we studied the lowest radial excitation of a degenerate Bose gas, which can be induced by an interaction quench. This excitation is also known as the breathing mode. Our theory shows an interesting phenomenon that the breathing mode frequency at unitarity returns to the value of a non-interacting gas. 
\begin{acknowledgments}
The authors thank Q. Zhou and F. Robicheaux for helpful discussions, and M. Eiles for careful reading of the manuscript. This work is supported by National Science Foundation and US-Israel Binational Science Foundation.
\end{acknowledgments}

\bibliography{apsrenormbec.bib}

\end{document}